# Dual-Phase MoS$_2$ And MXene Nanohybrids for Efficient Electrocatalytic Hydrogen Evolution

**Sichen Wei,[1] Yu Fu,[1] Huamin Li,[2*] and Fei Yao[1*]**

*1 Department of Materials Design and Innovation, University at Buffalo, the State University of New York, Buffalo, NY 14260, USA*

*2 Department of Electrical Engineering, University at Buffalo, the State University of New York, Buffalo, NY 14260, USA*

---

[†] Corresponding authors: huaminli@buffalo.edu, feiyao@buffalo.edu

## ABSTRACT

Molybdenum Disulfide ($MoS_2$) has been recognized as a potential substitution of Platinum (Pt) for electrochemical hydrogen evolution reaction (HER). However, the broad adoption of $MoS_2$ is hindered by its limited number of active sites and relatively low inherent electrical conductivity. In this work, we demonstrated a synergistic enhancement of both active site exposure and electrical conductivity by a one-step solvothermal synthesis technique. The 1T-phase enriched $MoS_2$ was directly formed on the titanium carbide ($Ti_3C_2Tx$, MXene) with carbon nanotubes (CNTs) acting as crosslinks. The existence of edge-enriched metallic phase $MoS_2$, the conductive backbone of MXene along with the crosslinking function of CNTs clearly improved the overall electrical conductivity of the catalyst. Moreover, the integration of two-dimensional (2D) $MoS_2$ with MXene effectively suppressed the MXene oxidation and 2D layer restacking, leading to good catalytic stability. As a result, an overpotential of 169 mV and a low Tafel slope of 51 mV/dec was successfully achieved. This work provides a new route for 2D-based electrocatalyst engineering and sheds light on the development of the next-generation PGM-free HER electrocatalysts.

**INTRODUCTION**

Hydrogen shows great potential in reducing greenhouse gas emissions and improving energy efficiency due to its environmentally friendly nature and inherent high gravimetric energy density [1–3]. Hydrogen gas can be generated via electrochemical water splitting based on the hydrogen evolution reaction (HER, $2H^+ + 2e^- \rightarrow H_2$) [4, 5]. HER is a multistep reaction that starts with a Volmer step ($H^+ + e^- + * \rightarrow H_{ads}$). The intermediate $H_{ads}$ is removed from the catalyst surface either by Tafel reaction ($H_{ads} + H_{ads} \rightarrow H_2 + 2*$) or by Heyrovsky reaction ($H^+ + H_{ads} + e^- \rightarrow H_2 + *$) [6–9]. The reaction kinetics is greatly affected by the number of available active sites (represented as * in the equation), the way $H_{ads}$ interacting with the catalyst surface (*i.e.*, hydrogen adsorption energy $\Delta G_{ads}$), and electron transfer rate. It is well known that Pt-group metals (PGM) are excellent catalysts for HER, but their practical applications are limited by the high cost and scarcity [1,7–11]. Therefore, the development of active HER catalysts made from low-cost materials is a key step in the utilization of hydrogen energy.

Among various PGM-free catalysts, two-dimensional (2D) molybdenum disulfide ($MoS_2$) is regarded as a promising alternative of Pt due to its earth abundance and near-zero $\Delta G_{ads}$ [3,12–14], and their properties are determined by its polymorph types, namely hexagonal 2H or trigonal 1T phases. The intrinsically low electrical conductivity arising from the semiconducting nature of the 2H phase hinders the development of $MoS_2$-based electrocatalysts. Tremendous efforts have been devoted to improving the conductivity of 2H-$MoS_2$ via inducing phase transition from semiconducting 2H phase to metallic 1T phase through vacancy control, doping engineering, and strain application, etc. [15, 35–37]. Among them, solution-phase-based ion intercalation achieved during $MoS_2$ hydrothermal/solvothermal synthesis has been considered as one of the most effective approaches for phase transition. Compared to multistep electrochemical/chemical intercalations where post-synthetic treatment is normally required, such a process enables ion intercalation along with $MoS_2$ synthesis in a single step one-pot reaction and produces highly active, defect rich 1T $MoS_2$ with high surface area, offering significant advantages such as simplicity, low cost, scalability, operational stability, and environmental friendliness due to the use of benign precursors/solvents. Compared with a hydrothermal synthesis which is widely used for producing various $MoS_2$ structures including nanosheets, nanodots, and nanoflowers with different 1T phase ratios, solvothermal synthesis involving the use of bisolvent is rarely explored. The use of bisolvent will increase the solubility of $MoS_2$ precursors, trigger deep 2H to 1T phase transition and prevent undesired oxidation reaction which is of particular importance for sensitive materials, such as MXene.

MXene ($M_{n+1}X_nT_x$, where "M" represents early transition metal, "X" is carbon and/or nitrogen, and $T_x$ is surface functional groups in the chemical formula) is a group of materials consisting of transition metal carbides/nitrides/carbonitrides, which are produced by selectively etching of "A" element from MAX phase ("A" is group IIIA and IVA element )[40,41]. It has been widely investigated in supercapacitors and aqueous batteries because of its high conductivity, excellent hydrophilicity, and large interlayer distance [42]. Compared with charge-neutral graphene, MXene exhibits a negatively charged surface due to the existence of enriched surface functional groups, including -OH, -O, and -F, *etc* [43,44]. These surface functional groups will not only enhance the dispersion of precursor but also promote the $MoS_2$ nucleation in bisolvent, making it a superior substrate for the synthesis of $MoS_2$. Especially, oxygen-containing functional groups have been proved with near-zero $\Delta G_{ads}$, suggesting potentially high catalytic activity for HER [45]. Recently, 2H $MoS_2$ has been integrated with different types of MXene using in-situ sulfidation and microwave-assisted growth methods. The hybrid structures delivered an exceptional catalytic activity and durability for HER. Nevertheless, the above-mentioned strategies normally require multiple engineering processes involving complicated experimental setups that are costly and time-consuming. Hydrothermal methods have also been employed to synthesize 2H $MoS_2$/MXene composites. But the low electrical conductivity of 2H phase $MoS_2$ is still a limiting factor for the overall catalytic performance.

Herein, we employ a one-step solvothermal method to synthesize dual-phase $MoS_2$ with an enriched 1T phase content and $Ti_3C_2T_x$ MXene composites using bisolvent as catalysts for HER. To further improve the catalytic activity, carbon nanotubes (CNT) are introduced in the hybrid structure as crosslinks. The ternary composite not only improves the overall electrical conductivity but also prevents the undesired oxidation and restacking of 2D materials simultaneously. Moreover, we notice that 1T phase-enriched $MoS_2$ nanoflakes are vertically integrated with MXene with good uniformity, leading to the exposure of numerous edge sites which are catalytic active for HER. As a result, an improved HER activity was achieved compared to pure 2H $MoS_2$ and other binary counterparts. The ternary composite exhibits an overpotential of 169 mV which is achieved at a current density of 10 mA/cm$^2$ and a low Tafel slope of 51 mV/dec with good stability [46,47].

**METHODS**

**Materials.** Anhydrous ethanol, lithium fluoride (LiF), hydrochloric acid (HCl), ammonium molybdate, thiourea and N, N-dimethylformamide (DMF) were purchased from Fisher Scientific, USA. $Ti_3AlC_2$ was purchased from Beijing Forsman Scientific Co. Ltd., China. Multiwalled carbon nanotubes (CNT) were purchased from XFNANO, China. All chemicals were used as received without any further purification.

**Preparation of Ti₃C₂Tx.** The Ti$_3$C$_2$ was prepared by the "MILD" etching method. Specifically, 4 g LiF powder was slowly added into 80 ml HCl (9 M) solution and stirred for 30 minutes until the LiF was fully dissolved. 4 g Ti$_3$AlC$_2$ was slowly added into the LiF/HCl mixture which was placed in an ice bath subsequently. The solution was kept at 40 °C for 48 hours with continuous stirring. After the reaction, the black powder was collected by centrifugation and washed with DI water until the supernatant reached a pH value of 6. The powder was dried under vacuum at 60 °C for 12 hours.

**Synthesis of the MoS₂-MXene-CNT electrocatalyst.** 1 g of as-prepared Ti$_3$C$_2$ powder and 0.1g CNT were added into 60 mL DI water/DMF (volume ratio 1:1) solvent, followed by ultrasonication to form a homogenous suspension. 1.928 g ammonium molybdate and 3.645g of thiourea were slowly added into the suspension and stirred until the precursors were well mixed. The suspension was then transferred into a 100 mL Teflon-lined autoclave and kept at 195 °C for 22 hours. After naturally cooling down to room temperature, the product was collected by centrifugation and washed with DI/ethanol, followed by vacuum drying at 60 °C for overnight. The as-prepared sample was denoted as MTC100 where the number indicated the weight of CNT (100 mg). Different binary composites (MoS$_2$-Ti$_3$C$_2$, MoS$_2$-CNT) and MTC hybrid structures with different MXene to CNT ratios were also synthesized similarly as control samples.

**Characterization.** X-ray diffraction (XRD) was performed using a Rigaku Ultima IV with Cu Kα radiation (wavelength = 1.541 nm). Raman spectra were collected using Renishaw InVia with an excitation laser wavelength of 514 nm. The morphologies of all products were investigated by field-emission scanning electron microscope (FE-SEM, Carl Zeiss AURIGA CrossBeam with Oxford energy dispersive x-ray spectra (EDS) system). The transmission electron microscopy (TEM) was conducted using the JEM ARM 200F system. X-ray photoelectron spectroscopy (XPS) was performed using a monochromatic Al Kα source (hv = 1486.6 eV) (ESCALAB 250, Thermo Scientific). The Brunauer-Emmett-Teller (BET) measurements were conducted on a Micromeritics Tri-Star II system by nitrogen ($N_2$) adsorption-desorption isotherm at 77 K.

**Electrochemical Measurements.** The ink for the HER test was prepared by dissolving 10 mg of as-prepared powder in a mixture of 500 µL of ethanol, 500 µL of DI water, and 15 µL of Nafion D-521 solution. The electrochemical characterization was performed using CHI760E electrochemical workstation (CH Instrument) in a standard three-electrode system which consists of a silver/silver chloride (Ag/AgCl in 1M KCl), a platinum (Pt) wire, and an ink-coated glassy carbon rotating ring disc electrode as reference, counter and working electrodes, respectively. The loading amount of the sample is 0.285 mg/cm$^2$ and the samples were cycled 20 times before any data recording. Nitrogen gas saturated 0.5 M $H_2SO_4$ was employed as electrolyte. All the measured potentials were converted to the potential vs. $E_{RHE}$ based on the equation: $E_{RHE} = E_{Ag/AgCl} + 0.059$ $pH + 0.222$. The linear sweep voltammetry (LSV) was carried out at a scan rate of 5 mV/s and the built-in IR compensation was executed prior to LSV tests. The electrochemical impedance spectroscopy (EIS, Biologic) was conducted from 0.1 Hz to 1M Hz with an amplitude of 5 mV at an overpotential of 250 mV vs. RHE. The electrochemical surface area was tested by cyclic voltammetry (CV) in the potential range of 0.05 ~ 0.15 V vs. RHE with different scan rates (20, 40, 50, 60, 80, and 100 mV/s). The double-layer capacitance ($C_{dl}$) was assessed from the slope of the linear regression between the current density differences ($\Delta J/2=(J_{anode}-J_{cathode})/2$ at an overpotential of 0.1 V vs. RHE) versus the scan rates. The accessible surface area of as-synthesized samples could be approximated from the electrochemical active surface area (ECSA). The ECSA was determined by ECSA=$C_{dl}/C_s$, where $C_s$ stands for the specific capacitance of standard electrode materials on a unit surface area. Here, based on the literature reported $C_s$ values for carbon electrode materials, 0.04 mF/cm$^2$ was used for ECSA calculations **[48]**. CV was performed between -0.3 V and 0.2 V vs. RHE to check the cycle stability.

**RESULTS AND DISCUSSION**

The MoS$_2$/Ti$_3$C$_2$ MXene/CNT samples were prepared by a one-step bisolvent solvothermal synthesis technique and the schematic illustration of the preparation procedure can be found in **Figure 1a.** Briefly, few-layered Ti$_3$C$_2$ MXene flakes were prepared by the "MILD" etching method. The mixture of MXene and CNT powders were then added into DI water/ethanol bisolvent along with ammonium molybdate and thiourea which served as Mo and S sources, respectively. The use of bisolvent is beneficial for increasing the solubility of MoS$_2$ precursors and therefore promoting MoS$_2$ nucleation. In addition, compared with the pure aqueous-based hydrothermal method, the adoption of bisolvent will preserve the high conductivity of the MXene matrix by improving the stability of MXene via reducing the undesired oxidation reaction. As shown in **Figure S1**, a large portion of TiO$_2$ nanoparticles was observed in pure aqueous-based synthesized MXene due to severe oxidation.

To confirm the structure and composition of the products, a series of characterizations were conducted. The SEM images of the as-synthesized materials are shown in **Figure 2**. The morphologies of dual-phase MoS$_2$ (DP-MoS$_2$) nanosheets and MXene layers are shown in **Figures 2a** and **b**, respectively. Compared to the pure DP-MoS$_2$ flakes which exhibit a well-defined nanoflower-like structure with the tendency to form aggregated bundles, 1T-enriched MoS$_2$ flakes tend to grow in the interlayers as well as the surface of Ti$_3$C$_2$ MXene, as shown in **Figure 2c**. Such a sandwiched DP-MoS$_2$/MXene structure will not only prevent the 2D layers from restacking but also protect the environmentally sensitive MXene from oxidation. The morphology of the DP-

MoS$_2$/ MXene/CNT composite (denoted as MTC100 hereafter) can be found in **Figure 2d**, where the three components are clearly observed.

**Figure 2e** displays the TEM image of MTC100 where a ternary composite consists of 1D/2D hybrid structures can be clearly observed. Although the density of CNTs is low, they served as a crosslink between the 2D islands and led to the formation of a well-connected conductive network, leading to an improved electron transfer efficiency within the whole system. The EDS elemental mapping can be found in the bottom panel of **Figure 2** (**Figure 2h-l**) where a homogeneous distribution of Mo, S, Ti, and C elements are revealed. The interlayer distance of DP-MoS$_2$ in the as-synthesized ternary composite was extracted from the HRTEM image (**Figure 2f**) and the value is ~ 0.98 nm which is clearly larger than the pure 2H-MoS$_2$ obtained from in situ sulfurization method (~ 0.61 nm). Besides that, the Ti$_3$C$_2$ MXene layers also exhibit an expanded interlayer spacing of ~ 1.5 nm (**Figure S2**) compared to the typical reported value of 1.1 nm **[53-55]**. The increased interlayer space can be ascribed to the NH$_4$ ions intercalation and the integrated stacking nature of the 2D layers. Moreover, we noticed that a large portion of DP-MoS$_2$ flakes was in favor of exposing edge sites (**Figure 2g**) in the MTC100 composite, leading to the involvement of numerous catalytic active sites for the future HER test.

**Figure 3a** shows the XRD patterns of various as-prepared samples. The DP-MoS$_2$ obtained via bisolvent synthesis exhibits major diffraction peaks at 9, 33.2, and 58.6 degrees, which are corresponding to the (002), (100), and (110) planes, respectively. Compared with the 2H-MoS$_2$ obtained in aqueous solution (see **Figure S3**), the (002) peak has downshifted from ~14 degree to 9 degree, suggesting an expanded interlayer spacing. To analyze in detail, the interlayer spacing has been extracted using Bragg's diffraction equation, and the value is calculated to be 0.98 nm for DP-MoS$_2$, which is 0.34 nm larger than the reported value of 2H-MoS$_2$ (0.64 nm). A similar peak shift has been observed for DP-MoS$_2$/CNT composite as well. Interestingly, this difference is very close to the size of NH$_4^+$ ion (0.35 nm), indicating the expanded interlayer spacing of DP-MoS$_2$ could be attributed to the interaction of (NH$_4$)$^+$, as indicated in **Figures 1d and e**. In our bisolvent synthesis process, both ammonium molybdate and DMF can act as the abundant source of (NH$_4$)$^+$ **[59, 60]**. The diffraction pattern of Ti$_3$C$_2$ MXene shows obvious peaks at 9, 18.4, 27.7, and 60.5 degrees, corresponding to (002), (006), (008), and (110) crystal planes. The absence of the Al peak at ~ 38 degree suggests the successful etching of the Al layer from the MAX precursor **[43,61,62]**. Two peaks located at 35.9 and 41.7 degrees come from TiC, which is the impurity in the MAX precursors **[63,64]**. In the case of the DP-MoS$_2$/Ti$_3$C$_2$ composite, two peaks located at 6.6 and 8.9 degrees can be clearly identified, corresponding to the (002) planes of Ti$_3$C$_2$ and MoS$_2$ with interlayer distances of 0.99 nm and 1.5 nm, respectively **[65]**. This result proves that the interlayer spacing of both MoS$_2$ and Ti$_3$C$_2$ were expanded simultaneously by forming the MoS$_2$/Ti$_3$C$_2$ composite. Similar interlayer distance expansion is observed in MTC100, which corresponds with previous TEM observations in **Figure 2f**. However, compared with the DP-MoS$_2$/Ti$_3$C$_2$, no further (002) peak shift is observed in MTC100, indicating that the addition of CNT did not impact the interlayer distance of MoS$_2$ and Ti$_3$C$_2$.

The Raman spectra of selected samples are shown in **Figure 3b**. The as-synthesized DP-MoS$_2$ sample shows the characteristic A$_{1g}$ and E$_{2g}$ vibration peaks at 378 cm$^{-1}$ and 405 cm$^{-1}$, respectively, corresponding to the 2H phase characteristics. In addition to the 2H peaks, three 1T-MoS$_2$ related peaks (150, 195, and 336 cm$^{-1}$) are observed which can be ascribed to J1, J2, and J3 vibration modes, respectively (marked in red, blue, and green in **Figure 3c)**. The coexistence of 1T and 2H related peaks in Raman again proved that the successful preparation of 1T enriched-MoS$_2$ using the bisolvent synthesis approach **[15,66-69]**. The formation of the 1T phase is triggered by the ion intercalation. Specifically, the intercalated $(NH_4)^+$ could stimulate the charge imbalance between $Mo^{3+}$ and $Mo^{4+}$ and cause the S plane sliding and the $MoS_2$ crystal structure distortion along with expanded interlayer spacing, resulting in the phase transformation from 2H to 1T eventually, as shown in **Figure 1e** and **f**. In the pure $Ti_3C_2$ Raman spectrum, several peaks that appeared at the lower Raman shift region (below 1000 cm$^{-1}$) can be attributed to the C-Ti vibrations, and the peaks located at 1341cm$^{-1}$ and 1583cm$^{-1}$ reflect the carbon-based D band and G band vibrations, respectively. In the case CNTs, the peak located at ~1356 cm$^{-1}$, 1584 cm$^{-1}$, and 2695 cm$^{-1}$ can be attributed to the vibration modes of D, G, and G' bands, respectively. The MTC100 sample exhibits both 2H and 1T phase $MoS_2$ related peaks along with CNT-related vibration peaks, indicating the successful formation of DP-$MoS_2$/MXene/CNT ternary composite. Moreover, the relative intensity of $A_{1g}$ to $E_{2g}$ peaks in **Figure 3c** provides insightful information related to the $MoS_2$ planes. The out-of-plane $A_{1g}$ vibration mode in $MoS_2$ is preferentially excited for edge-terminated structure due to the polarization dependence **[28,70]**. The $A_{1g}/E_{2g}$ ratio is extracted to be as large as 2, indicating an edge-enrich $MoS_2$ structure. This observation is corresponding to the TEM result shown in **Figure 2g**.

To investigate the chemical composition of the as-synthesized MTC100, XPS analysis was performed. The complete survey spectrum is shown in **Figure 4a** where the characteristic peaks of C, O, Mo, S, and Ti can be clearly observed. To analyze the details for each element, peak deconvolution was carried out. As shown in **Figure 4b**, four major peaks located at 284.1 eV, 284.85 eV, 285.85 eV and 287.3 eV for C1s spectra can be ascribed to the C-Mo/C-Ti, C-C, C-O, and C=C bonds, respectively. Relatively high intensity of C-Mo/C-Ti peak at 284.1eV may come from the C in multiple resources such as $Ti_3C_2$ and CNT which bonded with Mo and Ti. The peaks at 530.9 eV and 531.6 eV in the O1s spectrum (**Figure 4c**) originate from the C-Ti-(OH) and C-

Ti-O, respectively, corresponding to the –(OH) and -O functional group attached on the $Ti_3C_2$ surface. The peak located at 530.1eV can be assigned to the Ti-O bond, indicating the partial oxidation of the $Ti_3C_2$ MXene surface during the XPS measurement. Moreover, two peaks at 532.3 eV and 532.8 eV are related to $Al_2O_3$ and adhesive water, respectively. As shown in **Figure 4d**, the peaks for $Mo^{4+}$(d3/2) and $Mo^{4+}$(d5/2) are located at 231.8 eV and 228. 6eV, respectively, with an $S^{2+}$ peak at 225.8 eV. It is noteworthy that a doublet (at 229.6 eV and 233.1 eV) for $Mo^{5+}$ were observed, which is beneficial for the $MoS_2$ stability **[50]**. Besides that, $Mo^{6+}$ shows a peak at 235.7 eV, suggesting partial oxidation of Mo. As for S 2p spectra, two doublets that emerged at 163.5/162.1 eV and 162.7/161.4 eV can be attributed to 2H-$MoS_2$ and 1T-$MoS_2$, respectively, demonstrating the successful synthesis of mixed 1T and 2H phase $MoS_2$. Our calculation based on S 2p peak deconvolution suggested that the content of 1T phase $MoS_2$ is around 66%. **Figure 4f** displays the Ti 2p spectra. The doublet located at 464.3 eV and 458.9eV can be assigned to the Ti-O bond. The two peaks at 465.1 eV and 459.4 eV represent the Ti-F bond, which is formed during the Al layer etching process. The Ti-C bond-related peaks are observed at 461.2 eV and 456.6 eV, respectively.

The electrochemical HER activity exploration of as-synthesized $MoS_2$-based electrocatalysts was conducted in 0.5 M $H_2SO_4$ using a standard three-electrode configuration with the built-in IR compensation function. The HER performance of aqueous synthesized $MoS_2$ (2H-$MoS_2$) and DMF/water bislovent synthesized $MoS_2$ (DP-$MoS_2$) were compared. The linear sweep voltammetry (LSV) profile is displayed in **Figure S4a**. Compared to 2H-$MoS_2$ with an overpotential of 309 mV at a current density of 10 mA/cm$^2$ ($\eta$10), the DP-$MoS_2$ showed an obviously reduced overpotential (238 mV) as well as an improved Tafel slope (85mV/dec for DP-$MoS_2$ and 134 mV/dec for 2H-$MoS_2$, respectively). This enhanced HER performance can be ascribed to the high conductivity of metallic 1T-phase, which is also proved by the significantly reduced semicircle of 1T-enriched $MoS_2$ from the Nyquist plot in **Figure S4b**.

We then characterized the superior EC performance of as-prepared MTC100, as shown in **Figure 5**. **Figure 5a** displays the LSV curves of as-prepared samples. The pristine CNT and $Ti_3C_2$ (marked with the purple and red line, respectively) show nearly inert HER activity with a horizontal LSV curve pattern, indicating that CNT and $Ti_3C_2$ have limited contribution toward HER activity. The DP-$MoS_2$ possesses a much lower $\eta 10$, as shown with the black line in **Figure 5a**. The improved performance of DP-$MoS_2$ can be attributed to the metallic nature of 1T phase $MoS_2$ which significantly reduces the barrier for charge transport. Among all the samples, MTC100 exhibits the lowest $\eta 10$ of 169 mV compared to the other binary composites, such as DP-$MoS_2$/CNT (202 mV) and DP-$MoS_2$/$Ti_3C_2$ (230 mV), which highlights the crosslinking function of CNTs for a universical conductive network formation.

To further understand the kinetics behind HER, the Tafel plot is extracted to investigate the rate-determining step. In the case of pure DP-$MoS_2$ and pure $Ti_3C_2$, the values are 85 and 179 mV/dec, respectively. In the case of MTC100, the Tafel slope is extracted to be 51 mV/dec, indicating a Heyrovsky reaction governed reaction kinetics. The value is much lower than that of DP-$MoS_2$ binary composites (102 mV/dec for DP-$MoS_2$/$Ti_3C_2$ and 72 mV/dec for DP-$MoS_2$/CNT, respectively). Electrochemical impedance spectroscopy was conducted to investigate the origin of the better performance in the ternary hybrid structure. As shown in **Figure 5c,** the MTC100 shows the smallest charge transfer resistance ($R_{ct}$), which can be attributed to the boosted overall electrical conductivity by connecting different 2D islands via the introduction of CNT. Moreover, the $R_{ct}$ of DP-$MoS_2$/CNT is smaller than that of DP-$MoS_2$/$Ti_3C_2$, which may arise from the superior conductivity of CNT compared to $Ti_3C_2$.

The CV was carried out to determine the double-layer capacitance ($C_{dl}$) and to calculate the electrochemical surface area (ECSA), as shown in **Figure S5**. The measured $C_{dl}$ for DP-MoS$_2$, Ti$_3$C$_2$, DP-MoS$_2$/Ti$_3$C$_2$, DP-MoS$_2$/CNT and MTC100 is 11.55, 70.58, 67.03, 54.14 and 87.94 mF/cm$^2$, and the corresponding ECSA are calculated to be 288.75, 1764.5, 1675.75, 1352.5 and 2198.5 cm$^2$, respectively. More details are displayed in **Table S1**. The result shows that MTC100 clearly possesses the largest ECSA, indicating the abundant active sites within the catalyst. The BET specific surface area was also tested by N$_2$ adsorption/desorption isotherm, as shown in **Figure S6**. The result shows that MTC100 contains the largest surface area of 31m$^2$/g, which is in agreement with the ECSA analysis. Turnover frequency (TOF) is another vital parameter to evaluate the activity of an HER electrocatalyst which characterizes the intrinsic activity of an electrocatalyst at a single active site. The TOF values of the above-mentioned samples are calculated based on the CV measurement in a pH-neutral phosphate buffer solution. The calculation detail can be referred to the Method session. The TOF from an overpotential of 0.2V to 0.23V was plotted in **Figure 5d**. Among all the samples, MTC100 exhibits a superior TOF compared to other DP-MoS$_2$-based binary composites.

To investigate the cycling stability, the CV cycles between 0.3 V and -0.2 V were performed and the polarization curves before and after cycles are shown in **Figure 5e**. A neglectable shift of the polarization curve after 1000 CV cycles was observed, suggesting a long lifetime of MTC100 in the acidic media. The minimal activity loss can be attributed to the slow conversion of 1T enriched-MoS$_2$ to 2H-MoS$_2$ due to the metastable nature of 1T phase MoS$_2$. **Figure 5e** displays the time-dependent stability of MTC100. The current density was measured at an overpotential of 169 mV, which is based on the LSV data from **Figure 5a**, and the current density was stabilized at near 10 mA/cm$^2$ for more than 25 hours, further implies the outstanding stability of MTC100.

Based on the above characterizations and electrochemical measurements, the significant enhancement of MTC HER activity could be attributed to the following aspects. First, the MXene flakes provide a template for $MoS_2$ to grow vertically, which ensures the maximum exposure of active $MoS_2$ edge sites. The reduced oxidation of MXene because of the use of bisolvent synthesis along with the high coverage of $MoS_2$ also ensured the good conductivity of the composite. Second, the creation of DP-$MoS_2$/$Ti_3C_2$ mixture expanded the interlayer spacing and prevented the restacking of $MoS_2$ and $Ti_3C_2$ simultaneously, which promotes the contact between electrolyte and $MoS_2$, leading to an increased hydrogen ion adsorption. Third, the coexistence of 1T and 2H phase $MoS_2$, $Ti_3C_2$, and CNT provides overall high electrical conductivity, minimizing the charge transfer resistance during the reaction.

To optimize the HER activity, we synthesized four different DP-$MoS_2$/MXene/CNT samples with different weights of CNTs (50, 70, 100, and 200 mg). As shown in **Figure 6a**, the MTC100 possesses the smallest overpotential (169 mV). Moreover, the Tafel plot shown in **Figure 6b** proved that MTC100 exhibits a small Tafel slope of 51 mV/dec. It is noteworthy that MTC200 possesses an even smaller Tafel slope (49 mV/dec) than MTC100, which can be ascribed to the higher ratio of CNT that further enhanced the conductivity. The Nyquist plots are shown in **Figure 6c** and the EIS results reveal a remarkably reduced charger transfer resistance of MTC200, followed by MTC100, MTC70, and MTC50. It is worth mentioning that the small charge-transfer resistance of MTC200 could be from the high conductivity of CNT, which is well correlated to its small Tafel slope. The TOF of MTC100 is also the largest among different CNT amounts, as shown in **Figure 6d**, indicating an enhanced HER activity. It has been proved that the number of active sites and the electrical conductivity are two vital factors for determining the HER activity of an electrocatalyst. In our case, by adding more CNT, the conductivity of the electrocatalyst improved significantly, as can be proved by the small $R_{ct}$ and Tafel slope. Yet the relatively high CNT ratio will decrease the content of DP-MoS$_2$, leading to a decreased total number of active sites and therefore the catalytic activity correspondingly. Overall, the MTC100 provides a well-balanced relationship between electrical conductivity and the number of active sites and therefore delivered the best catalytic performance for HER.

**CONCLUSION**

In summary, we have successfully synthesized a ternary, dual-phase MoS$_2$/MXene/CNT composite by a one-step solvothermal technique, and the HER activity was investigated. Based on the material characterizations and electrochemical measurement results, we have demonstrated that the ternary hybrid structure showed a superior HER activity due to the synergistically optimized active site exposure and electrical conductivity. Specifically, the existence of metallic phase MoS$_2$, the conductive backbone of MXene along with the crosslink function of CNTs is beneficial for the overall electrical conductivity of the catalyst. Other than that, the integration of MoS$_2$ with MXene effectively suppressed the MXene oxidation and 2D layer restacking, leading to good catalytic stability. As a result, the optimized MTC ternary composite exhibits a significant enhancement of HER activity with the overpotential of 169 mV and Tafel slope of 51 mV/dec, as well as the long-term cycle stability. Our work shed light on the development of the next-generation PGM-free HER electrocatalysts.

**ASSOCIATED CONTENT**

**Supporting Information**

**AUTHOR INFORMATION**

**Corresponding authors**

huaminli@buffalo.edu, feiyao@buffalo.edu

**Author contributions**

H. L. and F. Y. conceived and supervised the project. S. W. synthesized materials performed materials characterization and electrochemical measurement. All authors discussed the results and contributed to the final manuscript.

**Acknowledgments**

This work was partially supported by New York State Energy Research and Development Authority (NYSERDA) under Award 138126, and the New York State Center of Excellence in Materials Informatics (CMI) under Award C160186. The authors acknowledge support from the Vice President for Research and Economic Development (VPRED) at the University at Buffalo.

**Ethics declarations**

The authors declare no competing interests.

# FIGURES

## Table of Content

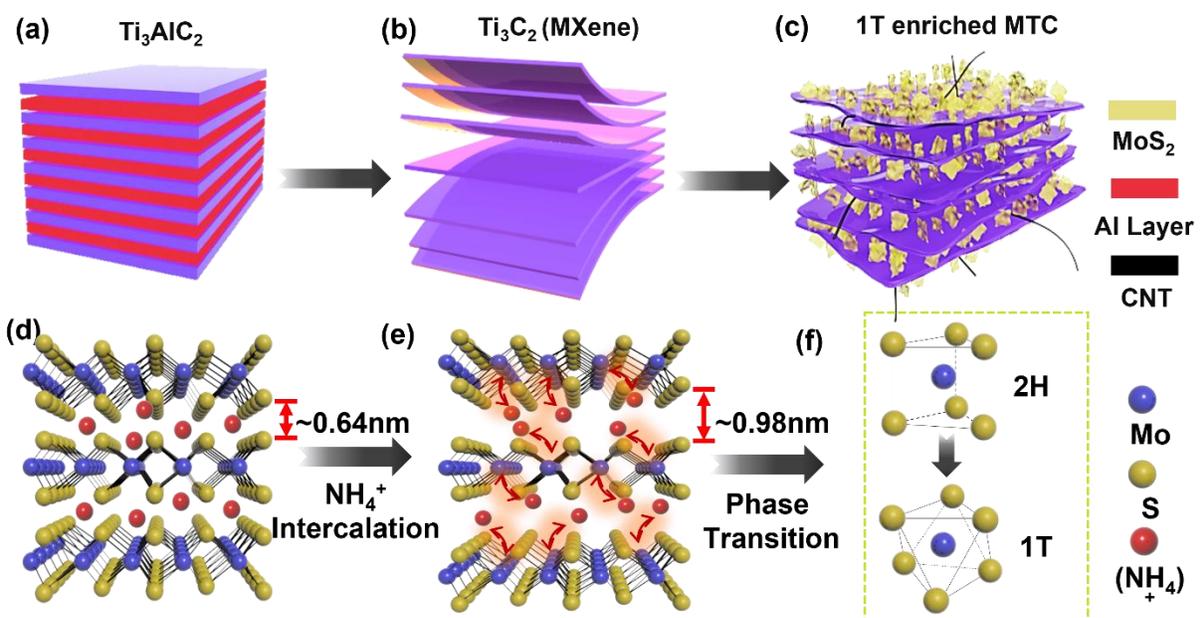

**Fig. 1**. Schematic illustrations of the preparation of 1T enriched-MoS$_2$/MXene/CNT composite through one-step solvothermal technique (a-c) and the mechanism of ammonia ion intercalation induced MoS$_2$ phase transition (d-f).

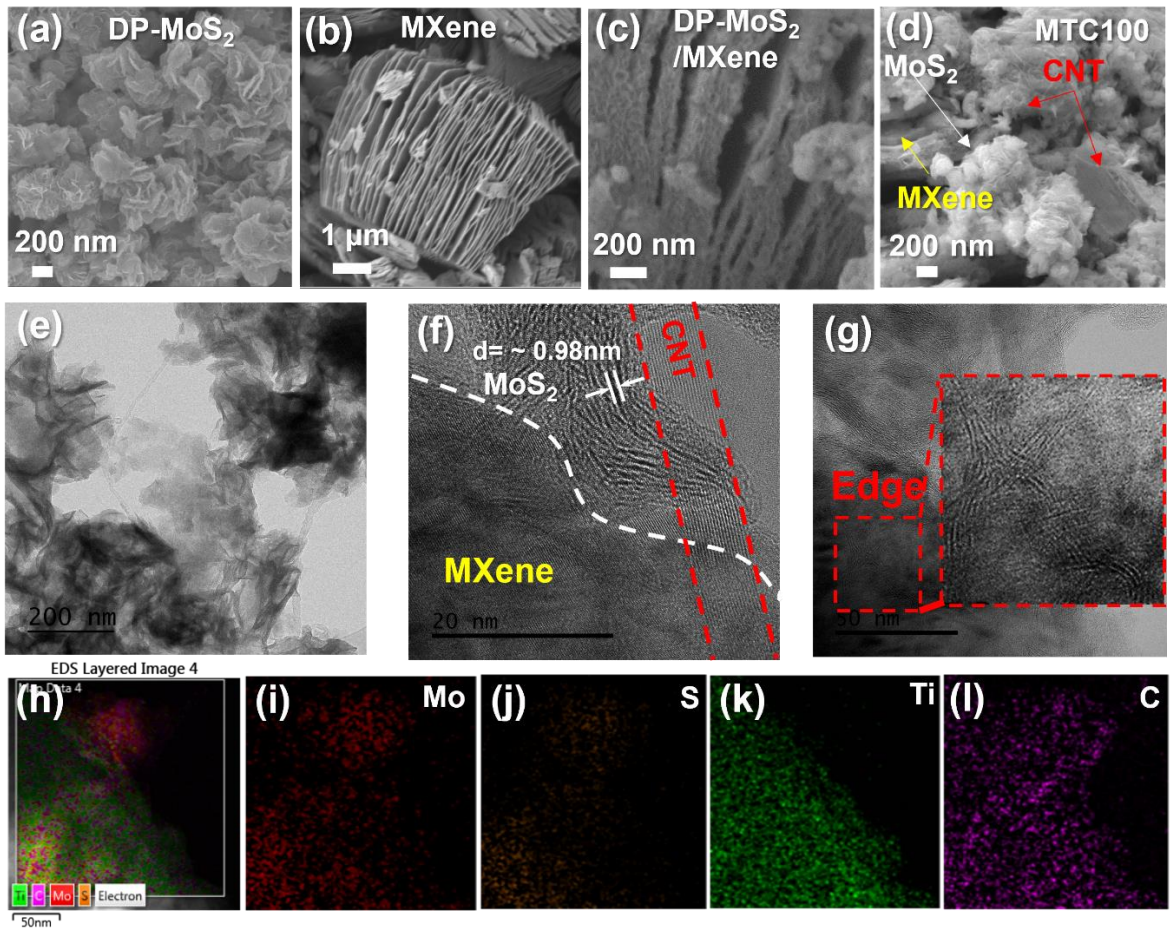

**Figure 2.** The morphology characterization of different samples. (a-d) are the SEM images of pure 1T enriched-MoS$_2$, MXene, MoS$_2$/Ti$_3$C$_2$T$_x$, and MTC100, respectively. (e) and (f) are the TEM image and HRTEM image of MTC composite. The corresponding EDS elemental mapping is shown in (h)~ (l).

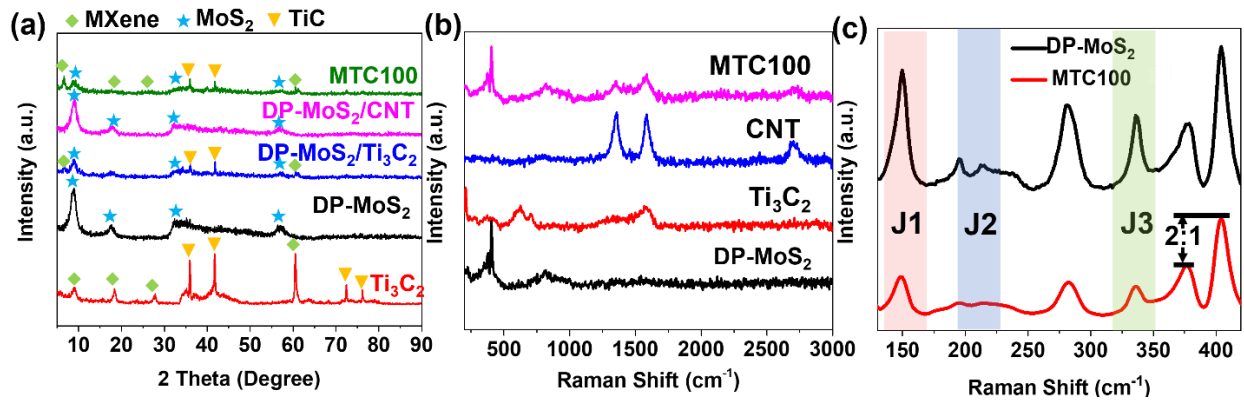

**Figure 3.** The XRD patterns (a) and the Raman spectra (b-c) of pristine DP-MoS$_2$, Ti$_3$C$_2$, CNT, and different composites.

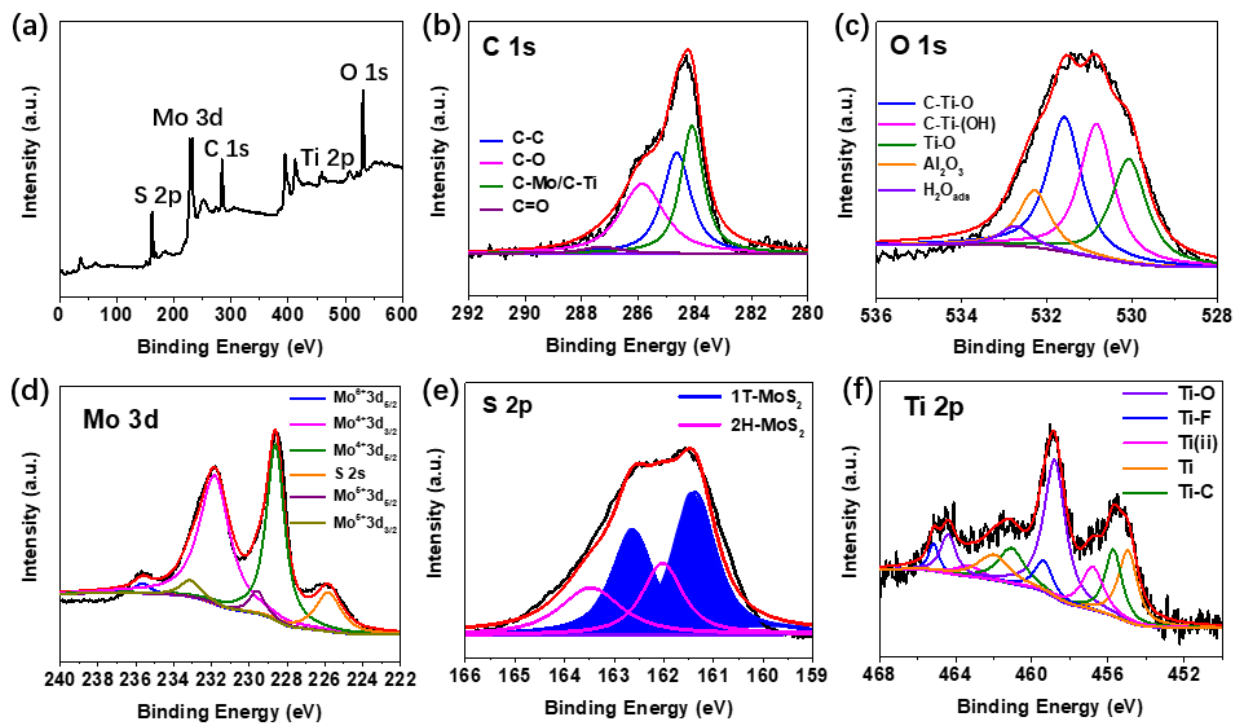

**Figure 4.** (a) The complete survey of XPS spectra. (b-f) The XPS spectra of MTC100 showing the binding energy of carbon, oxygen, molybdenum, sulfur, and titanium.

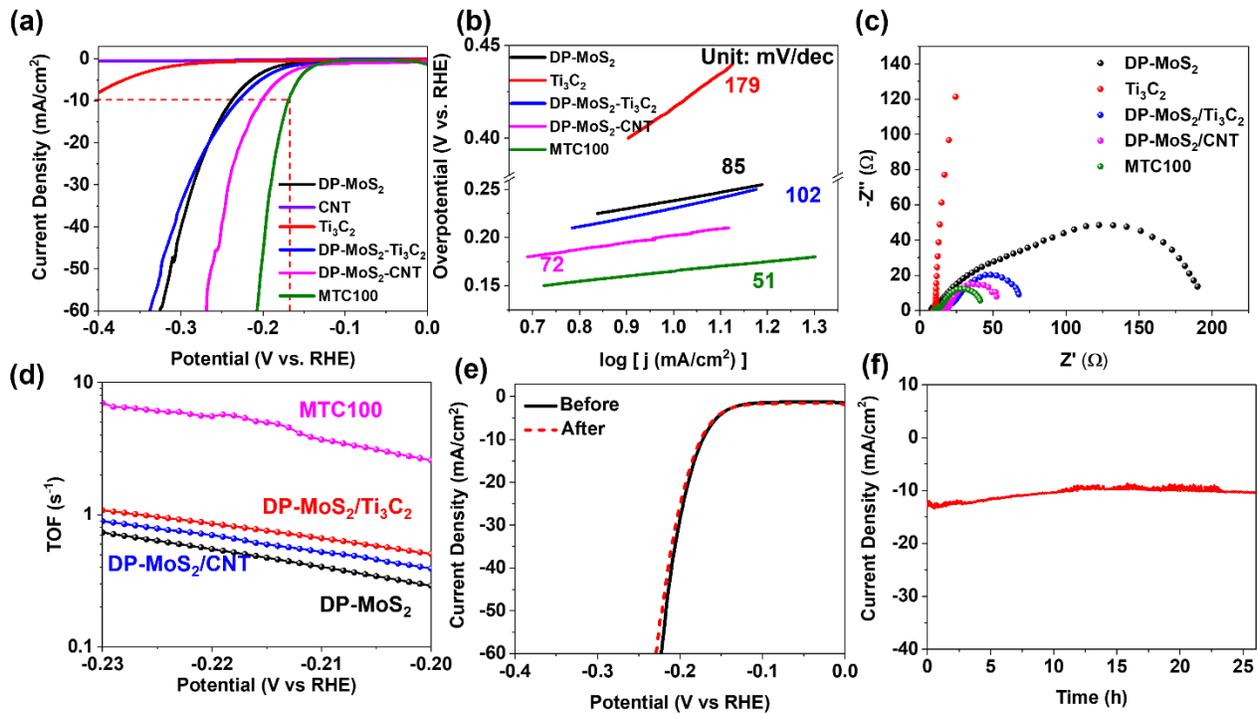

**Figure 5.** (a) Polarization curves measured at a scan rate of 5 mV/s and (b) Tafel plots for selected samples. (c) The Nyquist plot of different samples. (d) The turnover frequency versus potential plot. (e) The polarization curves of MTC1001 before and after 1000 cycles of CV scan and (f) the time-dependent stability plot, demonstrate its cycling stability.

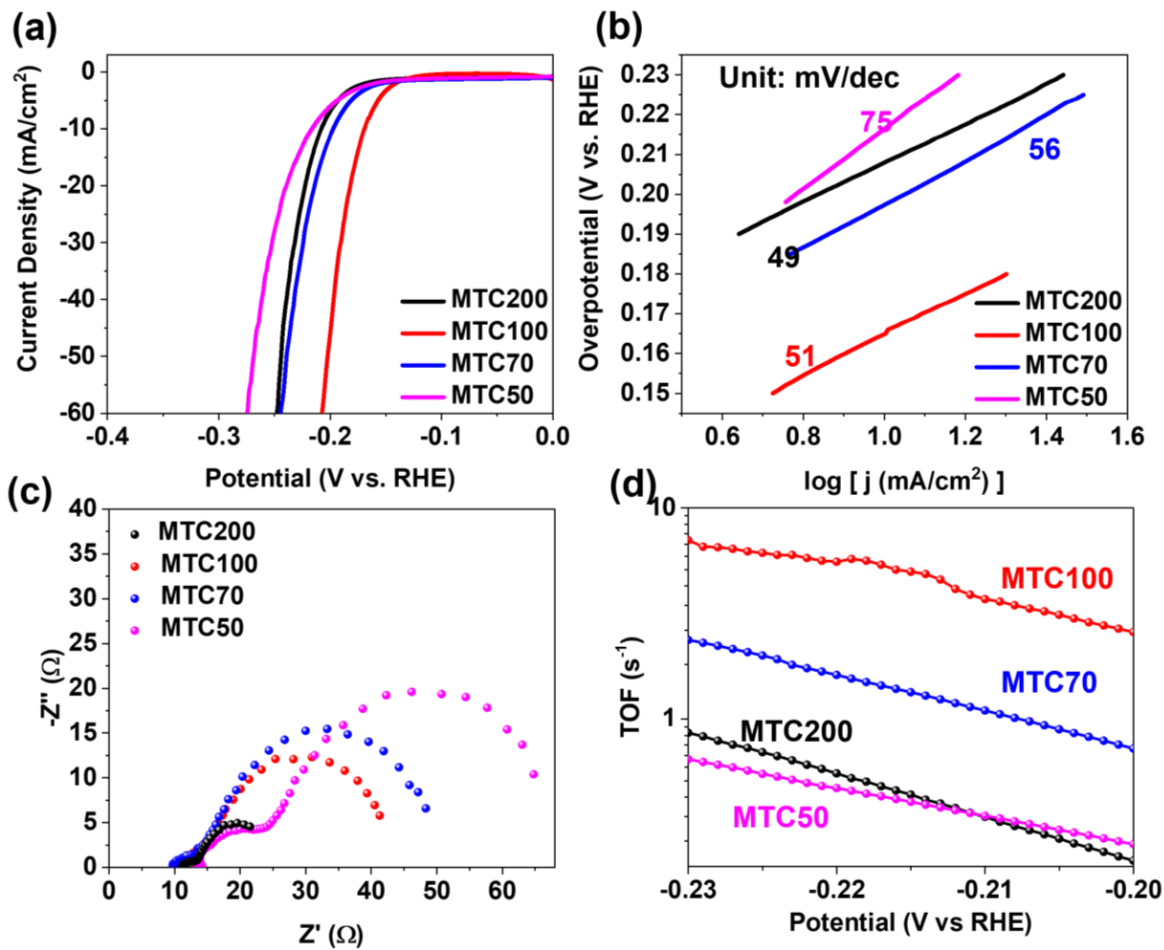

**Figure 6.** (a) Polarization curves, (b) Tafel plots, (c) Nyquist plots, and (d) turnover frequency versus potential plots of different samples.